\documentclass[journal]{IEEEtran}
\pdfoutput=1
\usepackage{cite}
\usepackage{pgf}

\usepackage{mathtools}
\usepackage{amssymb}
\usepackage{bm}

 \def\cF{{\mathcal{F}}}  
   
 \def\cN{{\mathcal{N}}}

 \def\cZ{{\mathcal{Z}}} 

\def\ba{{\mathbf{a}}}     
 \def\bff{{\mathbf{f}}}  \def\bh{{\mathbf{h}}}
     
 \def\bn{{\mathbf{n}}}   
     
\def\bu{{\mathbf{u}}}  \def\bv{{\mathbf{v}}}   \def\bx{{\mathbf{x}}}
\def\by{{\mathbf{y}}}  \def\bz{{\mathbf{z}}}
  \def\bzero{\mathbf{0}}

\def\bA{{\mathbf{A}}} \def\bB{{\mathbf{B}}}  
 \def\bF{{\mathbf{F}}} \def\bG{{\mathbf{G}}} \def\bH{{\mathbf{H}}}
\def\bI{{\mathbf{I}}}   
   
\def\bQ{{\mathbf{Q}}} \def\bR{{\mathbf{R}}} \def\bS{{\mathbf{S}}} 
\def\bU{{\mathbf{U}}} \def\bV{{\mathbf{V}}} \def\bW{{\mathbf{W}}} 
 \def\bZ{{\mathbf{Z}}}

\DeclareMathOperator{\vectorize}{vec}

\DeclareMathOperator*{\argmin}{arg\,min}

\DeclareMathOperator{\Real}{Re}
\DeclareMathOperator{\Imag}{Im}
\DeclareMathOperator{\diag}{diag}

\DeclareMathOperator{\trace}{Tr}

\def\complexs{{\mathbb{C}}}
\def\expec{{\mathbb{E}}}
\newcommand{\Exp}[1]{\expec\left\{#1\right\}}

\newcommand{\dd}{\mathop{}\!\mathrm{d}}

\newcommand{\Tr}{\mathsf{T}}

\usepackage{microtype}
\usepackage[hidelinks]{hyperref}
\hypersetup{pdfauthor={Stephen G.\ Larew},
						pdftitle={Adaptive Beam Tracking with the Unscented Kalman Filter for Millimeter Wave Communication}}

\begin{document}

\title{Adaptive Beam Tracking with the Unscented Kalman Filter for Millimeter Wave Communication}
\author{Stephen~G.~Larew,~\IEEEmembership{Student~Member,~IEEE} and David~J.~Love,~\IEEEmembership{Fellow,~IEEE}%
\thanks{Stephen G.\ Larew was supported in part by the National Science Foundation Graduate Research Fellowship Program under Grant No.\ DGE-1333468.  The authors are with the \href{https://engineering.purdue.edu/ECE}{School of Electrical and Computer Engineering}, Purdue University, West Lafayette, IN 47907 USA e-mail: \href{mailto:sglarew@purdue.edu}{sglarew@purdue.edu}.}%
}

\maketitle

\begin{abstract}
	Millimeter wave (mmWave) communication links for 5G cellular technology require high beamforming gain to overcome channel impairments and achieve high throughput.  While much work has focused on estimating mmWave channels and designing beamforming schemes, the time dynamic nature of mmWave channels quickly renders estimates stale and increases sounding overhead.  We model the underlying time dynamic state space of mmWave channels and design sounding beamformers suitable for tracking in a Kalman filtering framework.  Given an initial channel estimate, filtering efficiently leads to refined estimates and allows forward prediction for higher sustained beamforming gain during data transmission.  From tracked prior channel estimates, adaptively chosen optimal and constrained suboptimal beams reduce sounding overhead while minimizing estimation error.
\end{abstract}


\section{Introduction}
\label{sec:intro}

\IEEEPARstart{M}{illimeter} wavelength communication systems stand as an important pillar for Fifth Generation (5G) cellular standards\cite{6824752}. Beyond the 6 GHz ceiling of current systems, the 20 to 100 GHz range of millimeter wave (mmWave) frequencies offers ample bandwidth for high throughputs but suffers from less favorable propagation conditions.  High antenna count arrays at the transmitter and receiver can provide large beamforming gains to overcome the greater path loss and shadowing.  The mmWave channel estimation and beamforming problem has been studied extensively\cite{6847111,6600706,6979963,7400949,7959169,7342886,6717211,7147840,7947217}, but only a handful of works\cite{7510902,7905941,8107474,7582545} have looked at mmWave channel estimation and beamforming with a temporal channel correlation model.

Suppose the mmWave multiple-input multiple-output (MIMO) channel that is to be estimated is parameterized by a set of state variables that evolve over time.
If the transmitter and receiver know the state variables, then due to the parameterization of the channel model, they have full channel state information (CSI).
With probabilistic information about a state space model of the channel, then statistical CSI is available.
If the state variables stochastically evolve over time according to such a known model, the channel can be predicted into the future. Moreover, optimal sounding beams can be chosen given the channel predictions.  Therefore, this work assumes the mmWave MIMO channel can be parameterized by a state space model and focuses on adaptively tracking channel state parameters over time.

Previous works\cite{7582545,7905941} identified the path gains and angles of arrival and departure as candidate parameters for a state space model of the mmWave channel.  While only the angles are modeled in\cite{7510902}, and\cite{7905941} adds path gains, both follow a simple zeroth-order motion model (i.e., innovations are entirely due to a noise process).  A first-order model with linear motion for angles is described in\cite{7582545} but the model is nonlinear and specific to the discrete lens array.  In this work, the proposed time evolution model for the state space parameters is linear and not tied to a specific mmWave array architecture.  Moreover, the model extends to higher-order motion models.

Although the state evolution model of this work is linear, the channel observation model is generally nonlinear.  For tracking channel state, the extended Kalman filter (EKF), as used in\cite{7905941,7510902}, approximates nonlinearities by making linear approximations around the current state. 
Specifically, the Jacobian matrices are substituted into the normal Kalman filter equations for the linear transformations.  The linearized transformations fail, however, when too much error propagates.  The conditions and assumptions allowing the linearization to proceed with minimal error are dependent on the current state estimate, the covariance magnitude, and the transformation, which may be time-varying\cite{Merwe:2004aa}.  Moreover, the Jacobian matrices are non-constant and generally must be derived, either analytically or computationally, as a function of the sounding beams.  Thus, as the system diverges from the original design conditions or the linearization weakens, the EKF offers uncertain performance with high computational overhead.

The unscented Kalman filter (UKF)\cite{1271397}
offers an alternative suboptimal approximation given the intractability of recursive Bayesian estimation and the weaknesses of the EKF.  The UKF adapts the standard Kalman filter framework, which is the optimal exact solution for linear Gaussian systems, with an alternative to the linearization of the EKF.  Key to the UKF, the unscented transform propagates means and covariances through nonlinearities with improved accuracy.
This work employs the UKF because it readily adapts to model changes\footnote{Consider extensions to Section~\ref{sec:channel_model} with higher order motion models and noise correlation.}, presents a black box approach that scales with the state dimensions, and is not specific to any mmWave hardware architecture.

\section{System Model and Problem Formulation}
\label{sec:modelproblem}

\subsection{Millimeter Wave Communication Model}

Consider transmission and reception with \(M_{\mathrm{T}}\) and \(M_{\mathrm{R}}\) antennas, respectively.  The transmitter scales a symbol \(s_k\in\complexs\)
with the beamformer \(\bff_k\in\complexs^{M_\mathrm{T}}\) before transmission over the MIMO channel \(\bH_k\).  At the receiver, 
the combiner \(\bz_k\in\complexs^{M_\mathrm{R}}\) scales the noisy received signal
before the channel outputs are summed to form the received sample \(r_k\).  From input to output, the \(i\)th beamformed and combined noisy channel output during the \(k\)th channel coherence period is
\begin{align}\label{eq:io}
	r_k[i] = \bz_k^H[i] \big( \bH_k \bff_k[i] s_k[i] + \bn_k[i] \big) .
\end{align}
The noise \(\bn_k[i] \sim \mathcal{CN}(\bzero,\frac{1}{\rho}\bI_{M_{\mathrm{R}}})\) is independent and identically distributed across space and time.  The beams, a nonspecific term referring to either a transmit beamformer or receive combiner, are unit-norm constrained, i.e., \(\| \bff_k \|_2 = \| \bz_k \|_2 = 1\).

To estimate CSI, the transmitter and receiver sound beams during a channel coherence period.  For each of the \(N_\mathrm{T}\) beams in the columns of \(\bF_k\in\complexs^{M_\mathrm{T}\times N_\mathrm{T}}\), the receiver sounds the beams in the columns of \(\bZ_k\in\complexs^{M_\mathrm{R}\times N_\mathrm{R}}\).  During the \(k\)th coherence period, the \(i=1,\ldots,N_\mathrm{R}N_\mathrm{T}\) channel  observations \(r_k[i]\) form a noisy version of the matrix \(\bZ_k^H \bH_k \bF_k\).

\subsection{Millimeter Wave Channel Model for Tracking}
\label{sec:channel_model}

Standard multipath channel models used for lower frequencies have been adapted to mmWave frequencies\cite{7400949}.
In the following, a time dynamic state space model is built up from the common mmWave channel model.

The mmWave channel is commonly modeled with array steering vectors for the arriving and departing plane waves.  
Consider the one-dimensional uniform linear array (ULA)
of \(M\) antennas with antenna separation \(d\), for which the array steering vector is
\begin{IEEEeqnarray}{c}
	\ba(\nu) = \begin{bmatrix} e^{-j2\pi \nu} &\cdots & e^{-j2\pi m\nu} & \cdots & e^{-j2\pi M\nu} \end{bmatrix}^T .
	\label{eq:array_steering_vector}
\end{IEEEeqnarray}
The angle of arrival \(\theta\in[0,2\pi)\) relates to the normalized spatial angle
\(\nu=\frac{d}{\lambda}\sin\theta\) (\(\lambda\) is wavelength).
The \(\ell\)th of \(L\)  paired angle of departures (AOD) and angle of arrivals (AOA) from the transmitter to the receiver has gain \(\alpha_\ell\in\complexs\) and normalized spatial AOA \(\nu_{\mathrm{R},\ell}\) and AOD \(\nu_{\mathrm{T},\ell}\). The narrowband slowly varying MIMO channel at time \(k\) is\cite{7400949}
\begin{IEEEeqnarray}{rCl}
	\bH_k &=& \sum_{\ell=1}^{L} \alpha_{k,\ell} \ba\left(\nu_{\mathrm{R},k,\ell}\right)\ba^H\left(\nu_{\mathrm{T},k,\ell}\right) ,
	\label{eq:channel_model_static}
\end{IEEEeqnarray}
where the channel parameters \(\alpha_{k,\ell}\), \(\nu_{\mathrm{R},k,\ell}\), and \(\nu_{\mathrm{T},k,\ell}\) for \(\ell=1,\ldots,L\) fully describe a given channel instance at time \(k\) and suggest explicit or implicit inclusion in a state space model.

Each multipath between the transmitter and receiver is modeled by a virtual position and velocity, which together form the virtual state.
Consider an imaginary plane at a fixed distance from and parallel to an antenna array.  The true angle and angular velocity of the multipath, the source of the plane wave, are projected onto the imaginary plane to create the virtual state.  Without loss of generality, the virtual plane is one unit away from the ULA so that the virtual position and velocity of a path are measured with the same arbitrary unit.
The virtual position \(\upsilon\) is related to its angle \(\theta\) by \(	\upsilon = \tan(\theta) \).
Thus, its normalized spatial angle \(\nu\) for (\ref{eq:array_steering_vector}) is
\begin{IEEEeqnarray}{rCl}
	\nu &=& \frac{d}{\lambda}\sin\left( \arctan\left( \upsilon \right) \right)
	\\
	&=& \frac{d}{\lambda}\frac{\upsilon}{\sqrt{1+\upsilon^2}} .
	\label{eq:normalized_spatial_angle_from_virtual_position}
\end{IEEEeqnarray}

For a single channel realization (temporarily dropping the time subscript), consider the state space for the virtual positions \(\upsilon_{a,\ell}\) and gains \(\alpha_\ell\) of the \(\ell=1,\ldots,L\) multipaths in relation to the transmitter (\(a=\mathrm{T}\)) and receiver (\(a=\mathrm{R}\)).  As a first order approximation to the dynamics of the multipaths\footnote{Higher order models may prove more accurate or yield better performance, but we adopt a linear model for simplicity, brevity, and clarity of exposition.}, also consider the first time derivative \(\dot{\upsilon}_{a,b}=\frac{\dd}{\dd t} \upsilon_{a,b}\) of the virtual positions.
The complete state space that parameterizes the channel is
\begin{IEEEeqnarray}{c}
	\bx = \vectorize
	\begin{bmatrix}
		\bx^{(\alpha)} & \bx^{(\upsilon_\mathrm{T})} & \bx^{(\upsilon_\mathrm{R})}
	\end{bmatrix} ,
	\label{eq:multipath_state_space}
\end{IEEEeqnarray}
where the state vectors are
\begin{IEEEeqnarray}{rCl}
	\bx^{(\alpha)} &=&
	\vectorize
	\begin{bmatrix}
		\Real\alpha_1 & \cdots & \Real\alpha_L
		\\
		\Imag\alpha_1 & \cdots & \Imag\alpha_L
	\end{bmatrix} ,
	\\
	\bx^{(\upsilon_a)} &=&
	\vectorize
	\begin{bmatrix}
		\upsilon_{a,1} & \cdots & \upsilon_{a,L} &
		\\
		\dot{\upsilon}_{a,1} & \cdots & \dot{\upsilon}_{a,L}
	\end{bmatrix} , a \in \{\mathrm{T},\mathrm{R}\} .
	\label{eq:multipath_state_space_sub}
\end{IEEEeqnarray}

The macro level time dynamics of \(\bx^{(\upsilon_a)}\) are modeled as a first-order linear approximation of the multipath motion.
From time step \(k-1\), the virtual position state \(\bx_{k-1}^{(\upsilon_a)},a=\mathrm{T},\mathrm{R}\) is deterministically advanced at the macro level to the next state \(\bx_k^{(\upsilon_a)}\) by the matrix transformation
\begin{IEEEeqnarray}{c}
	\bA_k^{(\upsilon)} = 
  \bI_{L} \otimes
	\begin{bmatrix}
		1 & T_\mathrm{S} \\
		0 & 1
	\end{bmatrix} .
\end{IEEEeqnarray}
A single path gain is temporally correlated through a first order Gauss-Markov process, which gives the deterministic time step
\begin{IEEEeqnarray}{c}
	\bA_k^{(\alpha)} = \beta_k \bI_{2L} .
\end{IEEEeqnarray}
The overall discrete-time dynamic model for time step \(T_\mathrm{S}\) is
\begin{IEEEeqnarray}{rCl}
	\bx_k &=& \bA_k \bx_{k-1} + \bu_k ,
  \label{eq:channel_dynamics_model}
\end{IEEEeqnarray}
where \(\bu_k\) is process noise and the deterministic time step matrices are along the diagonal of \(\bA_k\), i.e.,
\begin{IEEEeqnarray}{rCl}
	\bA_k = \diag\left( \bA_k^{(\alpha)}, \bA_k^{(\upsilon)}, \bA_k^{(\upsilon)} \right) .
\end{IEEEeqnarray}

The micro level motion of the multipath is modeled by the process noise vector \(\bu_k \sim \mathcal{N}\left(\bzero, \bQ_k \right)\).
The upper left \(2L\times 2L\) matrix \(\bQ_k^{(\alpha)}\) of \(\bQ_k\) is the process noise covariance for the path gains.  Each path gain is independent of all others and follows a first order Gauss-Markov property giving
\begin{IEEEeqnarray}{c}
	\bQ_k^{(\alpha)} = \diag\left( \frac{1-\beta_1^2}{2}, \frac{1-\beta_1^2}{2}, \ldots, \frac{1-\beta_L^2}{2}, \frac{1-\beta_L^2}{2} \right) .
\end{IEEEeqnarray}
When \(\bQ_k^{(\alpha)}\) is diagonal\footnote{Advanced models with correlation in the process noise are ignored for simplicity, although they pose no problem for the tracking in Section~\ref{sec:ukf}.}, the dynamic model for \(\bx^{(\alpha)}\) in (\ref{eq:channel_dynamics_model}) is simply an uncorrelated Rayleigh fading model with \(\beta_k\) relating to the Doppler spread of the channel\cite{61437}.
As for the virtual positions, the process noise for each position and derivative pair is a diagonal \(2\times 2\) matrix \(\bQ_k^{(\upsilon)}\).

\subsection{Channel Observation Model}

For the channel tracking problem, the observation model (\ref{eq:io}) is transformed from a complex to a real number representation to match the real number space used in the state evolution model.  Vectorizing the sounding outputs with a property of the Kronecker product gives
\begin{IEEEeqnarray}{c}\label{eq:io_sounding}
	\vectorize \left[ \bZ_k^H \bH\left(\bx_k\right) \bF_k \right] = \left( \bF_k^\Tr \otimes \bZ_k^H \right) \vectorize\left[ \bH\left(\bx_k\right) \right] ,
	\label{eq:observation_kronecker}
\end{IEEEeqnarray}
where \(\bH(\bx_k)=\bH_k\) is~(\ref{eq:channel_model_static}) evaluated with the state in \(\bx_k\).
Defining the observation matrix \(\bG_k = \bF_k^\Tr \otimes \bZ_k^H\) and the vector channel \(\bh(\bx_k) = \vectorize \bH(\bx_k)\), the noisy channel observation for the \(k\)th state is
\begin{IEEEeqnarray}{rCl}
	\by_k &= \bG_k \bh( \bx_k ) + \bv_k ,
	\label{eq:simple_observation_model}
\end{IEEEeqnarray}
where \(\bv_k \sim \mathcal{CN} \left( \bzero, \bI \right)\) is additive white Gaussian observation noise that is independent from the process noise.

To transform the complex observations (\ref{eq:simple_observation_model}) to the real domain,
first, the real and imaginary components of the vectorized complex channel are separately stacked to give the real-valued channel
\begin{IEEEeqnarray}{rCl}
	\tilde{\bh}(\bx_k) = \begin{bmatrix} \Real\bh(\bx_k) \\ \Imag\bh(\bx_k) \end{bmatrix} .
	\label{eq:real_channel_model_vectorized}
\end{IEEEeqnarray}
The complex-valued matrix transform in (\ref{eq:simple_observation_model}) becomes
\begin{IEEEeqnarray}{rCl}
	\tilde{\bG}_k =
	\begin{bmatrix}
		\Real\bG_k & -\Imag\bG_k \\
		\Imag\bG_k & \Real\bG_k
	\end{bmatrix} ,
\end{IEEEeqnarray}
so the real-valued observation is
(\(\tilde{\bv}_k \sim \cN ( \bzero, {1}/{2}/{\rho} \bI ) \))
\begin{IEEEeqnarray}{rCl}
	\tilde{\by}_k &= \tilde{\bG}_k \tilde{\bh}( \bx_k ) + \tilde{\bv}_k .
	\label{eq:real_observation_model}
\end{IEEEeqnarray}

\subsection{Problem Description}
\label{sec:probdesc}

Design an observation matrix \(\bG_k = \bF_k^\Tr \otimes \bZ_k^H \) and track the current channel state \(\bx_k\) of the system from the observations \(\by_{1:k}=\left\{ \by_i : i=1,\ldots,k \right\}\).  The two problems are as follows.
\begin{enumerate}
	\item First, track the current channel state \(\bx_k\) from \(\by_{1:k}\).
	\item Secondly, design good sounding beams for tracking with low overhead.
\end{enumerate}

\section{Adaptive Channel Tracking}

To improve the communication system performance and reduce estimation overhead, this section develops a tracking method for the channel parameters given an initial estimate.  The system model is a channel model (\ref{eq:channel_model_static}) that is a nonlinear function of a state space (\ref{eq:multipath_state_space}) that evolves over time according to (\ref{eq:channel_dynamics_model}).  Together, (\ref{eq:channel_dynamics_model}) and (\ref{eq:real_observation_model}) form a discrete-time dynamic system.

\subsection{Optimal Bayesian Estimation}

The optimal Bayesian estimator of the current channel state \(\bx_k\) from the observations \(\by_{1:k}\) involves recursively updating the posterior density \(p( \bx_k | \by_{1:k} )\), allowing the minimum mean square estimator \(\hat{\bx}_k = \Exp{ \bx_k | \by_{1:k} }\) to be derived.  However, finding the densities or the MMSE estimator is generally only tractable for linear Gaussian systems, in which case the Kalman filter is the exact optimal recursive solution.  For the model in Section~\ref{sec:modelproblem} and many others, the nonlinear and non-Gaussian properties necessitate approximate solutions.

\subsection{Suboptimal Estimation with UKF}
\label{sec:ukf}

In this work, the UKF is employed for state estimation and tracking.
Key to the UKF is the unscented transform, which estimates the posterior distribution of a transformed random variable by propagating mean and covariance of the prior through the nonlinear function by way of \emph{sigma points}\cite{1271397}.
See\cite{6003762,1271397,Merwe:2004aa} for full treatments of the UKF.

First, an unspecified channel estimator gives the initial state estimate \(\hat{\bx}_0\) and its covariance \(\bR_0\).
Then, given the previous state estimate \(\hat{\bx}_{k-1}\) and its covariance \(\bR_{k-1}\), the state in the \(k\)th block is predicted to be
\begin{IEEEeqnarray}{c}
	\hat{\bx}_{k|k-1} = \bA_k \hat{\bx}_{k-1} 
\end{IEEEeqnarray}
with \emph{a priori} covariance
\begin{IEEEeqnarray}{rCl}
	\bR_{k|k-1} &=& \expec\left[ \hat{\bx}_{k|k-1}\hat{\bx}_{k|k-1}^\Tr \right]
	\\
	&=& \bA_k\bR_{k-1}\bA_k^\Tr + \bQ_k .
\end{IEEEeqnarray}

The statistics of observation (\ref{eq:real_channel_model_vectorized}) are approximated with the symmetric set\cite{7042740} of sigma points \(\left\{\bm{\chi}_i \right\}_{i=0}^{12L}\) that approximate \(\hat{\bx}_{k|k-1},\bR_{k|k-1}\).
Let \(\bm{\chi}_0 = \hat{\bx}_{k|k-1}\) and let the weights be \(\omega_0^{(m)} = \frac{\lambda}{6L+\lambda},\omega_0^{(c)} = \frac{\lambda}{6L+\lambda}+(1-\eta^2+\mu),\omega_k=\frac{1}{2(6L+\lambda)},k=1,\ldots,12L\) where 
\(\lambda=\eta^2(6L+\kappa)-6L\) (\(\eta\) and \(\kappa\) are described in\cite{Wan:2000aa}). 
The remaining sigma points are
\(	\bm{\chi}_i=\hat{\bx}_{k|k-1} \pm \left(\sqrt{(6L+\lambda) \bR_{k|k-1}}\right)_i \),
where \((\sqrt{\bR})_i\) denotes the \(i\)th column of the matrix square root of \(\bR\).
Transforming the sigma points through the channel gives \(\{\bm{\zeta}_i = \tilde{\bh}(\bm{\chi}_i) \}_{i=0}^{12L}\).
The (approximate) predicted mean and covariance of the real channel are, respectively,
\begin{IEEEeqnarray}{rCl}
	\hat{\bh}_k &=& \sum_{i=0}^{12L} \omega_i^{(m)} \bm{\zeta}_i
	\\
	\hat{\bm{\Pi}}_k &=& \sum_{i=0}^{12L} \omega_i^{(c)} \left( \bm{\zeta}_i - \hat{\bh} \right) \left( \bm{\zeta}_i - \hat{\bh} \right)^\Tr
\end{IEEEeqnarray}

The Kalman filter update follows.  The innovation and its (approximate) covariance are, respectively,
\begin{IEEEeqnarray}{rCl}
	\bar{\by}_k &=& \tilde{\by}_k - \tilde{\bG}_k \hat{\bh}_k
	\\
	\bS_k &=& \tilde{\bG}_k\hat{\bm{\Pi}}_k\tilde{\bG}_k^\Tr + {1}/({2\rho})\bI .
\end{IEEEeqnarray}
The (approximate) cross-covariance is
\begin{IEEEeqnarray}{rCl}
	\hat{\bR}_k^{(\bx\bh)} &=& \sum_{i=0}^{12L} \omega_i^{(c)} \left( \bm{\chi}_i - \hat{\bx}_{k|k-1} \right) \left( \bm{\zeta}_i - \hat{\bh}_k \right)^\Tr .
\end{IEEEeqnarray}
The updated state estimate and its covariance are, respectively,
\begin{IEEEeqnarray}{rCl}
	\hat{\bx}_k &=& \hat{\bx}_{k|k-1} + \hat{\bR}_k^{(\bx\bh)} \tilde{\bG}_k^\Tr \bS_k^{-1} \bar{\by}_k
	\\
	\bR_k &=& \bR_{k|k-1} - \hat{\bR}_k^{(\bx\bh)} \tilde{\bG}_k^\Tr \bS_k^{-1} \tilde{\bG}_k\hat{\bR}_k^{{(\bx\bh)}^\Tr} .
\end{IEEEeqnarray}

\subsection{Adaptive Channel Sounding}
\label{sec:adaptchansound}

This section shows how to adaptively design the sounding beams in \(\bF_k\) and \(\bZ_k\) at each sounding time \(k\) under the channel tracking method in Sec.~\ref{sec:ukf}.  For an objective function, between the mean squared error and average receive SNR metrics\cite{6777295}, minimizing the mean squared error \(\bR_k\) of the channel state is more general as it pursues overall accuracy of the channel estimate under the model.

The optimal beams constrained to the sets \(\cF\) and \(\cZ\) at the \(k\)th sounding period are
\begin{IEEEeqnarray}{rCl}
	\argmin_{\bF_k\in\cF,\bZ_k\in\cZ} &\quad& \trace \left( \bW \bR_k \right) , \label{eq:min_weighted_est_cov}
\end{IEEEeqnarray}
where \(\bW=\diag(w_1,\ldots,w_{6L})\) is a positive semi-definite weighting matrix.  
Ignoring the beam constraint sets and the Kronecker product constraint on \(\tilde{\bG}_k\), the optimal\footnote{Proof follows from extending the proof in\cite[Appendix~A]{6777295} for non-diagonal matrices and applying the properties of the  block generalized Rayleigh quotient\cite[p.~81-82]{Baker:2008aa}.} \(\tilde{\bG}_k^\Tr = \tilde{\bV}_{[1:N_\mathrm{T}N_\mathrm{R}]}\) where \(\tilde{\bV}\) are the ordered normalized generalized eigenvectors of \(\bA=\hat{\bR}_k^{{(\bx\bh)}^\Tr}\bW^{-1}\hat{\bR}_k^{(\bx\bh)}\) and \(\bB=\hat{\bm{\Pi}}_k+\frac{1}{2\rho}\bI\) (i.e., \(\bA\tilde{\bV}-\bB\tilde{\bV}\bm{\Lambda}=\bzero,\bm{\Lambda}=\diag(\lambda_1,\ldots,\lambda_{M_\mathrm{T}M_\mathrm{R}}),\lambda_1\ge\lambda_2\ge\cdots\)).  The beams in \(\bG\) are chosen to minimize the Frobenius norm of the difference to the optimal beams, giving
\begin{IEEEeqnarray}{rCl}
	\min_{\bF\in\cF,\bZ\in\cZ} &\quad& \left\| \bV_{[1:N_\mathrm{T}N_\mathrm{R}]} - \bF^* \otimes \bZ \right\|_F . \label{eq:opt_const_beams}
\end{IEEEeqnarray}
Optimizers for (\ref{eq:opt_const_beams}) generally depend on the constraint sets for sounding beams and hence the specific mmWave hardware.

Considering only a unit-norm constraint, the sounding beams follow from the best rank-one approximation of \(\bU\), which is a rearranged \(\bV_{[1:N_\mathrm{T}N_\mathrm{R}]}\) to match the terms when changing \(\bF^* \otimes \bZ\) to the outer product \(\vectorize(\bF^*)\vectorize(\bZ)^\Tr\).  Thus, (\ref{eq:opt_const_beams}) becomes
\begin{IEEEeqnarray}{rCl}
	\min_{\bF\in\cF,\bZ\in\cZ} &\quad& \left\| \bU - \vectorize\left(\bF^*\right) \vectorize\left(\bZ\right)^\Tr \right\|_F .
\end{IEEEeqnarray}
Optimized sounding beams in \(\vectorize\left(\bF^*\right)\) and \(\vectorize\left(\bZ\right)\) are the dominant left and right singular vectors of \(\bU\), with an additional normalization step for each beam.

\section{Simulation Results}
\label{sec:simulations}

The following simulations demonstrate the time dynamic channel model and channel tracking for a selected set of representative parameters.  Simulated time frames are 5 ms in which 50 channel observations are taken (\(T_\mathrm{S}=10^{-4}\)).  Between observations, state updates and predictions are made every 1 \(\mu\)s.  Channels have \(L=4\) paths independently and uniformly distributed across all AoD and AoA with the first derivative of the virtual position \(\dot{\upsilon}\) drawn from the Rayleigh distribution with variance \(\sigma_{\dot{\upsilon}}=100\sqrt{2/\pi}\) and multiplied by \(-1\) with probability one-half.
Other simulation parameters are virtual position estimates \(\hat{\upsilon}\sim\cN(\upsilon,0.1)\), velocities \(\hat{\dot{\upsilon}}\sim\cN(0,10^6)\), gain correlation \(\beta=0.905\), SNR is \(\rho=10\) dB, \(M_\mathrm{T}=M_\mathrm{R}=16\), and after refinement \(N_\mathrm{T}=N_\mathrm{R}=6\).  Sounding beams for observations are chosen as described in Sec.~\ref{sec:adaptchansound}.

\begin{figure}[!t]
\centering
\input{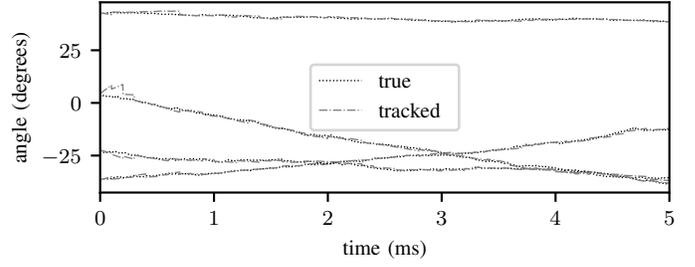}
\caption{Time evolution of paths and their tracked estimates.}
\label{fig:paths}
\end{figure}

Fig.~\ref{fig:paths} shows a representative example of the time evolution of the AoD of 4 paths.  To start, the tracked estimates suffer from the error in the initial estimate and lack of virtual position velocity.  By 1 ms, the tracked AoDs stabilize on the correct virtual positions and velocities are accurate.  From 2 to 5 ms, good tracking continues despite the crossing of paths.

\begin{figure}[!t]
\centering
\input{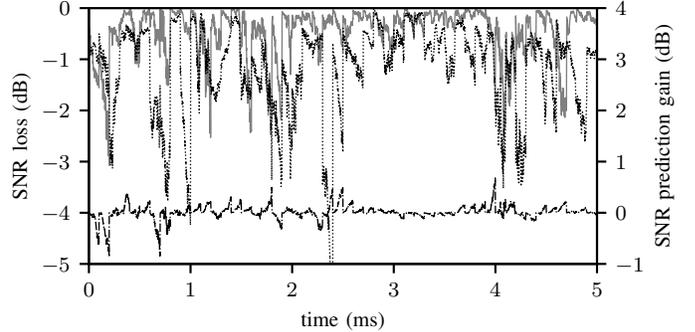}
\caption{(gray solid) Loss in SNR with tracked statistical CSI compared to perfect CSI.  (black dotted) Loss in SNR with one-shot CSI estimation compared to perfect CSI.  (dash dotted) Gain ratio of predicted to no-prediction SNR.}
\label{fig:esnr}
\end{figure}

Fig.~\ref{fig:esnr} shows the loss in effective SNR over time.  The loss is measured as the ratio of SNR from tracked statistical CSI to SNR with perfect CSI.  Also shown for comparison is the loss when a one-shot non-tracked estimate  with the same statistics as the initial estimate is taken every \(T_\mathrm{S}\) seconds.  Defining the SNR for a pair of beams as \(\Gamma_k(\bff,\bz) = \rho\left| \bz^H\bH_k\bff \right|^2 \), then the loss ratio at any 1 \(\mu\)s time period is \(\Gamma_k\left(\bff_\mathrm{opt},\bz_\mathrm{opt}\right) / \rho  \left\|\bH_k\right\|_2^2 \).  Optimal beams \(\bff_\mathrm{opt}, \bz_\mathrm{opt}\) are chosen as the dominant left and right singular vectors for the current tracked channel estimate.  First, note how SNR for the tracked statistical CSI generally improves on that for one shot estimation.  After 0.5 ms, the initial channel estimate is refined to improve the beamforming gain.  Losses in gain increase when paths cross (cf. Fig.~\ref{fig:paths}).
The improvement due to prediction between channel sounding is also shown.  Notably, when the loss increases significantly, state prediction can improve SNR.

\section{Conclusions}

A time dynamic channel model was proposed and both a recursive channel tracking algorithm (based on UKF) and adaptive sounding beam selection process were presented.  The channel model builds on static models by considering physical intuitions of the multipaths.  Moreover, the channel model as presented can be extended (e.g., higher order path motion) or modified considerably so long as a simple state space model can be derived.  
Channel tracking with optimal and constrained suboptimal beams allows long periods of time to pass between soundings.  Simulation results show promising tracking performance.  Future work may consider limited-feedback for CSI\cite{6784322,6777295} and more specific constrained beam sets reflecting hybrid analog and digital mmWave architectures.

\bibliographystyle{IEEEtran}
\bibliography{IEEEabrv,references}

\end{document}